\documentclass[12pt,a4paper]{article}
\usepackage[latin1]{inputenc}
\usepackage[T1]{fontenc}
\usepackage{amsmath}
\usepackage{amsfonts}
\usepackage[english]{babel}
\usepackage[dvips]{graphicx,color}
\begin{document}
\title{Functional approach to the\\ fermionic Casimir effect}
\author{C.~D.~Fosco and E.~L.~Losada\\
  {\normalsize\it Centro At\'omico Bariloche and Instituto Balseiro}\\
  {\normalsize\it Comisi\'on Nacional de Energ\'{\i}a At\'omica}\\
  {\normalsize\it R8402AGP S.~C.~de~Bariloche, Argentina.} }
\date{}
\maketitle
\begin{abstract}
We use a functional approach to calculate the Casimir energy due to Dirac
fields in interaction with thin, flat, parallel walls, which implement {\em
imperfect\/} bag-like boundary conditions. These are simulated by the
introduction of $\delta$-like interactions with the walls. We show that,
with a proper choice for the corresponding coupling constants,
bag-model boundary condition are properly implemented.

We obtain explicit expressions for the energies in $1+1$
and $3+1$ dimensions, for massless and massive fields.
\end{abstract}
\section{Introduction}\label{sec:intro}
In recent years, a renewed interest in the Casimir effect~\cite{rev}
emerged, as a consequence of new, refined experimental techniques~\cite{exp}, 
which raised the required standards of the theoretical calculations. 
They were usually based on gross simplifying assumptions, mostly about the
properties of the mirrors. 
To explain the experimental results, it is then important to 
use more accurate models, including corrections due, for example,
to the imperfect nature of the mirrors.

In this paper we are concerned with the calculation of the fermionic
Casimir effect for imperfect `mirrors', in a functional integral framework.
In a fermionic context, the mirrors are meant to partially reflect the
fermionic current; including as a  particular case the `bag' boundary
conditions, where the current is completely reflected.

From a technical point of view, we shall use an adapted version of a
previously used functional approach~\cite{funct}, to cope with the case of
fermionic fields and imperfect boundaries. In the fermionic context, this
method had already been applied to the case of fermions in $2+1$ dimensions
satisfying bag conditions on a curve and coupled to a gauge
field~\cite{Fosco:2006xg}.

To cope with imperfect mirrors, which will occupy parallel planes, we
follow~\cite{Sundberg:2003tc} to introduce a coupling of the bilinear
$\bar\psi \psi$ to a $\delta$-function potential of the proper coordinate.
We shall then calculate the resulting Casimir energy as a function of the
distance between two mirrors, using to that effect a functional formalism
which maps the original problem to a one-dimensional one, where the fields
live on the surfaces of the mirrors. We show that exact results may be
obtained by  for $1+1$ and $3+1$ dimensions; for the $1+1$ dimensional case
our result in the perfect bag case agrees with the one
of~\cite{Sundberg:2003tc}, in spite of the fact that the treatment of the
problem is quite different; in particular, the description of the
interaction cannot be mapped in a straightforward manner from one
calculation to the other.

This article is organized as follows: in section~\ref{sec:method} we
introduce the method in general; in~\ref{sec:casimir} we apply it to
the calculation of the Casimir energy for two imperfect mirrors in $1+1$
and $3+1$ spacetime dimensions. In section~\ref{sec:bag} we analyze the
relationship between this kind of interaction and the bag boundary
conditions, which emerge when the coupling constant takes a particular
value.  Finally, in section~\ref{sec:conc} we present our conclusions.
\section{The method}\label{sec:method}
In this section, we shall introduce the method used for the calculation of
the vacuum energy, in the presence of `defects' which generate
the approximate bag-like boundary conditions, for the case of a fermionic field in $D+1$
spacetime dimensions. 

Our approach is based on the well-known property that the vacuum energy,
$E_0$,  may be obtained from the Euclidean functional ${\mathcal Z}\equiv
e^{-\Gamma}$, as:
\begin{equation}\label{eq:defe0}
{\mathcal E}_0 \;=\; \lim_{T\to\infty} \frac{\Gamma}{T}\;,
\end{equation}
where $T$ is the extension of the (imaginary) time coordinate.
On the other hand, in the functional integral representation, ${\mathcal Z}$ 
is given, for a Dirac field,  by:
\begin{equation}
{\mathcal Z} \;=\; \int{\mathcal D}\psi {\mathcal D}\bar\psi \, e^{-S} ,
\end{equation}
 where $S$, the Euclidean action, will be such that $S\,=\,S_0\,+\,S_I$, with
\begin{eqnarray}
S_0 &=&\int d^{D+1}x \;{\bar\psi}(x)\; (\not \! \partial + m) \,\psi (x)\nonumber\\
S_I &=&\int d^{D+1}x \;{\bar\psi}(x)\; V(x) \,\psi (x) \;,
\end{eqnarray}
which are the free and interaction actions, respectively.

In the last equation, $V(x)$ denotes a `potential', introduced in order to
simulate the boundary conditions. Although we shall mostly consider just
the cases of one or two defects, we write (for the sake of generality) its
expression for a system with $N$ equally-spaced defects:
\begin{equation}
V(x) \; = \; g \, \sum_{\alpha = 0}^{N - 1} \delta(x_D - a_{\alpha})\; ,
\end{equation}
where $a_\alpha = \alpha \,L$ and $\alpha=0,1,\ldots,N-1$. $g$ is the
coupling constant, whose role in imposing an approximate boundary condition
is discussed in section~\ref{sec:bag}. It is worth noting here that, as
mentioned in~\cite{Sundberg:2003tc}, Dirac's equation with this kind of
potential is not well-defined. However, also in~\cite{Sundberg:2003tc}, a
consistent calculation of the Casimir energy for a Dirac field in $1+1$
dimensions could still be performed, since one can set up the problem in
terms that avoid the calculation of the eigenstates of the Dirac
Hamiltonian. We shall here use an approach that also bypasses the
eigenproblem for the Dirac Hamiltonian, relying instead on the
corresponding Euclidean propagator. The reason this can be done is, of
course, that the vacuum energy may be expressed as a function of the
expectation values of fermionic bilinears which are, at least in this case,
well-defined even when the eigensystem is not.

Of course, the total energy for this kind of potential will be, since there
is translation invariance along the coordinates parallel to the surface of
the mirrors, proportional to the `area' of each defect; thus we will be
interested in evaluating energy densities, or pressures, rather than the extensive
quantity $E_0$ (except the $D=1$ case where the two quantities coincide).  

In $D$ spatial dimensions, the defects are co-dimension $1$ hyperplanes. Putting
the system in a $d \equiv D-1$-dimensional `box' of side $a$, we obtain the
vacuum energy density ${\mathcal E}_0$ dividing  $E_0$ (calculated for such
a box) by $a^d$, and then taking the $a \to \infty$ limit: 
\begin{equation}\label{eq:defz}
{\mathcal E}_0 \;=\; \lim_{T,\, a\to \infty}\left(\frac{\Gamma}{a^d T}\right) \;.
\end{equation}
We have denoted by $x$ the coordinates in $D+1$ dimensional
spacetime $x=(x_0,\ldots,x_D)$.  It will turn out to be convenient to
introduce also coordinates $y\,\in\,{\mathbb R}^{d+1}$, for the subset of
the $d+1$ spacetime components that parametrize the $\alpha^{th}$ defect's
world-volume, as $x\, =\, (y,a_\alpha )$. 
Analogously, momentum-space coordinates in $D+1$ dimensions are denoted by
$p$, while $q$ is reserved for its $d+1$-dimensional restriction:
\begin{equation}
p\;=\;(q,p_D)\;,
\end{equation}
where $p_D$ is the momentum component along the normal direction to the
mirrors.

In terms of these conventions, we have: 
\begin{equation}
e^{-S_I} = \prod_{\alpha = 0}^{N - 1} e^{- \,g \, \int
d^{d+1}y\;{\bar\psi}(y, a_\alpha)\,\psi (y, a_\alpha)}
\end{equation}
which, following the methods of~\cite{funct},
may be written in an equivalent way by introducing a pair of auxiliary Grassmann fields
$\bar\chi_\alpha(y)$ and $\chi_\alpha(y)$ for each defect. The vacuum
functional then adopts the form:
\begin{eqnarray}\label{eq:Z2}
	{\mathcal Z} &=& {\mathcal Z}_0 \, \int \Big[\prod_{\alpha = 0}^{N - 1}
	\mathcal{D}\chi _\alpha\mathcal{D}\bar\chi _\alpha  \Big] \;\exp
	\Big[ \sum_{\alpha = 0}^{N - 1} 
	\int d^{d+1}y\;\chi^2_\alpha(y) \nonumber\\
	& + & g \, \int d^{D+1}x\int d^{D+1}x'\, \bar\eta (x)\,
	G^{(0)}(x,x')\,\eta(x')\Big]\,,
\end{eqnarray}
where:
\begin{equation}\label{def:eta}
 \eta(x) \,\equiv\, \;\sum_{\alpha = 0}^{N - 1}\chi_\alpha(y)\,\delta(x_d\,-\,a_\alpha)\,,
\end{equation}
while ${\mathcal Z}_0$ is the vacuum functional in the absence of defects: 
\begin{equation}
 {\mathcal Z}_0  \,=\, \int{\mathcal D}\psi {\mathcal D}\bar\psi \,
 e^{-S_0} \,=\, \det \big(\not \! \partial + m \big)\,.
\end{equation}
On the other hand, 
\begin{equation}
	G^{(0)}(x,x')\;=\; \langle x|(\not \! \partial + m)^{-1}| x'\rangle\,,
\end{equation}
is the free fermion propagator. Note that the auxiliary fields are
naturally defined on each one of the mirrors' world-volumes. 

Since we are interested in calculating the dependence of ${\mathcal E}_0$ on the
positions, $a_\alpha$, of the defects, we shall discard any constant which
is independent of those variables.  Constant factors in the functional
integral, like ${\mathcal Z}_0$, generate precisely that kind of constant, 
which we shall therefore ignore (without bothering to rename the corresponding
subtracted quantity).

Replacing (\ref{def:eta}) into (\ref{eq:Z2}), we may write the latter in
the more explicit form:
\begin{equation}\label{eq:ZparaN} 
\mathcal{Z} \,=\,\int \Big[\prod_{\alpha = 0}^{N - 1}\, \mathcal{D}\chi
_\alpha\mathcal{D}\bar\chi _\alpha  \Big] \; e^{\sum_{\alpha,\beta =
0}^{N - 1}\int d^{d+1}y\int d^{d+1}y'\, \bar\chi_\alpha (y)\, \mathcal{K}_{\alpha
\beta}(y,y')\,\chi_\beta(y')} \;, 
\end{equation}
where we introduced the matrix kernel:
\begin{align}\label{eq:KparaN} 
\mathcal{K}_{\alpha \beta}(y,y')\,=\,\delta_{\alpha\beta}\,\delta (y - y') + g\,G^{(0)}(y,a_\alpha,y',a_\beta)\,,
\end{align}
where each $\alpha, \, \beta$ element is also a matrix in Dirac space.

The formal result of evaluating the integral above is then:
\begin{equation}
 {\mathcal Z} \;=\; \det \Big[ \mathcal{K}_{\alpha \beta}(y,y') \Big]
\end{equation}
where the determinant is meant to affect all, continuum and discrete, variables. This means
that:
\begin{equation}
{\mathcal E}_0 \;=\; - \, \lim_{T,\, a\to \infty}
\left(\frac{1}{a^d T} {\rm Tr} \ln  {\mathcal K} \right) \;,
\end{equation}
where `${\rm Tr}$' is the trace over continuous and discrete
indices~\footnote{Again, a finite spacetime box is introduced before
tacking the limit.}.

Since there is invariance under translations in the $y$ coordinates, we may
Fourier transform with respect to them. The kernel becomes then block
diagonal in the continuous variables; besides, a proper counting of the
modes in the box shows that the $a^d T$ factor is cancelled by an identical one
coming from the numerator; thus, in the limit,
\begin{equation}\label{eq:e0gen} 
	{\mathcal E}_0 \;=\; - \int
	\frac{d^{d+1}q}{(2\pi)^{d+1}} \, {\rm tr} \big[\ln
	\widetilde{\mathcal K}(q) \big] \;, 
\end{equation}
where the tilde has been used to denote Fourier transformation, and `${\rm
tr}$' denotes trace over the $\alpha,\beta$ and Dirac indices only.

In the following section, we deal with the evaluation of the  previous
expression for ${\mathcal E}_0$ in the most relevant case, i.e., $N=2$.
\section{Casimir energy}\label{sec:casimir} Besides the subtraction of the vacuum energy in the absence of the defects,
there is another constant to get rid of: the self-energy of the defects. It
may be identified as the result of evaluating ${\mathcal E}_0$ for $L \to
\infty$.  Thus the prescrition for such a subtraction, which will render a
finite quantity as a result, is to subtract from ${\mathcal E}_0$ its limit
when $L \to \infty$. Of course, to give meaning to the expression at
intermediate steps, a regularization is introduced.

Since the last subtraction brings into play the self-energy of the defects,
we present first $N=1$, where the self-energy first emerges, and then the
$N=2$ case where we evaluate the energy for the case of two defects,
identifying and subtracting the contributions of the self-energies of the
two mirrors. 
\subsection{One defect}\label{ssec:one}
For $N = 1$, the $\widetilde{\mathcal K}$ matrix has only one element
($\alpha=\beta = 0$), thus there is no need to use $\alpha$, $\beta$ indices: 
\begin{equation}\label{eq:defw}
	\widetilde{\mathcal K} \,=\, 1 \,+\, \frac{g}{2} \,W(q)  \;,\;\;
W(q) \, \equiv \, \frac{m\,+\,i\not \! q}{\sqrt{m^2\,+\,q^2}}\;.
\end{equation}
so that ${\mathcal E}_0$ becomes:
\begin{equation}
	{\mathcal E}_0 \;=\; \int \frac{d^{d+1}q}{(2\pi)^{d+1}} \,
	{\rm tr} \Big\{\ln \big[ 1 \,+\, \frac{g}{2} W(q)\big]\Big\} \;.
\end{equation}

When $D=1$, the Euclidean gamma matrices have the single eigenvalues $\pm 1$. 
Then the vacuum persistence amplitude is:
\begin{equation}\label{eq:GammaN1d1}
	{\mathcal E}_0 \,=\,- \, \int \frac{dp_0}{2\pi} \, 
	\ln{\Big(}1+\frac{g^2}{4}+\frac{g\,m}{\sqrt{m^2 \,+\,p_0^2}}{\Big)}\;,
\end{equation}
which has a logarithmic UV divergence, easily regularized by using a
frequency cutoff.  

A lengthier, but quite straightforward calculation shows that, in $D = 3$:
\begin{equation}\label{eq:GammaN1d3}
	{\mathcal E}\,=\, -\;2\, \int \frac{d^3p}{(2\pi)^3} \, 
	\ln\left[1 + \frac{g^2}{4} + \frac{g\,m}{\sqrt{m^2 +
	q^2}}\right]\;,
\end{equation}
which is quadratically divergent. Again, an Euclidean can be used to give
meaning to the integral, whose explicit form we do not need: indeed, it
will emerge in the next subsection only to be subtracted in order to fix
the energy to zero when $L \to \infty$. 
\subsection{Two defects}\label{ssec:two}
When $N = 2$ the matrix $\widetilde{\mathcal K}(q)$ is given by
\begin{equation}\label{eq:k2}
\tilde{\mathcal K}(q)=\left[\begin{array}{cc}
      1 + \frac{g}{2} W(q) & \frac{g}{2}\Big(W(q) -\gamma _D \Big) 
     e^{-L\sqrt{m^2+q^2}} \\
                    \\
     \frac{g}{2}\Big( W(q) +\gamma _D \Big)e^{-L\sqrt{m^2+q^2}}& 
     1 + \frac{g}{2} W(q)  \\
\end{array}\right]
\end{equation}
with $W$ as defined in (\ref{eq:defw}).

Using the property: 
\begin{equation}\label{eq:baa} 
\det\,\widetilde{\mathcal K}(q)\,=\,\det \widetilde{\mathcal K}_{00}\;\det 
\widetilde{\mathcal K}_{11}\; \det\Big[ {\mathcal I}-\widetilde{\mathcal K}^{-1}_{00}
\widetilde{\mathcal K}_{01}\widetilde{\mathcal K}^{-1}_{11}
\widetilde{\mathcal K}_{10}\Big]
\end{equation}
for the determinant of the matrix in (\ref{eq:k2}), we may identify the
first two factors above as yielding the self-energies for the defects, as
studied in~\ref{ssec:one}. We first introduce an UV cutoff to give meaning
to those two factors (the third factor is convergent) before subtracting
the corresponding self-energies. The resulting pressure is:
\begin{equation}
	{\mathcal E}_0 \;=\; - \, \int \frac{d^{d+1}q}{(2\pi)^{d+1}} \,
	{\rm tr} \big[\ln  \widetilde{\mathcal M}(q) \big] \;,
\end{equation}
with 
\begin{equation} 
\widetilde{\mathcal M}(q)\,\equiv\, {\mathcal I}-\widetilde{\mathcal K}^{-1}_{00}
\widetilde{\mathcal K}_{01}\widetilde{\mathcal K}^{-1}_{11}
\widetilde{\mathcal K}_{10} \;.
\end{equation}

Let us now evaluate ${\mathcal E}_0$ above for $D=1$ and $D=3$.  In one
spatial dimension, if we take into account Dirac's indices,
$\widetilde{\mathcal K}_{\alpha\beta}$ is a matrix with two $2 \times 2$
blocks, while $\widetilde{M}$ is just a $2\times 2$ matrix.
A somewhat lengthy but otherwise straightforward calculation shows that the
latter has eigenvalues $\lambda_1, \,
\lambda_2$ given by: 
\begin{eqnarray}\label{eq:autK}
\lambda_1 &=& 1 \nonumber\\
\lambda_2 &=& 1
+\,\frac{16\,g^2\,\,q_0^2}{[(4\,+\,g^2)\,\omega(q_0)\,+\,4\,g\,m]^2}\,e^{-2L|q_0|}
\;,
\end{eqnarray}
where $\omega(q_0) \equiv \sqrt{q_0^2+m^2}$.
Therefore, the trace in the expression for the vacuum energy can be
calculated exactly. In the massless case, we can obtain the exact
expression for the energy as a function of $g$, since:
\begin{equation}{\label{eq:Casimir2Pd1} }
{\mathcal E}_0(L) = - \frac{1}{\pi}\;\int _0^{\infty}dq_0\;\ln \Big[1 +
\frac{16\,g^2}{(4+g^2)^2}\,e^{-2L\,q_0}\Big]\;,
\end{equation}
or,
\begin{equation}
	{\mathcal E}_0 \;=\; -\,\frac{c_1(g)}{L} 
\end{equation} 
where
\begin{equation}
c_1(g) \;\equiv\; - \frac{1}{2\,\pi }\,{\rm
	Li}_2\big[\frac{-16\,g^2}{(4 + g^2)^2}\big]\,,
\end{equation} 
is a dimensionless constant which determines the strength of the Casimir
interaction, and ${\rm Li}_n(x)$ denotes the polylogarithm function. In
figure (\ref{fg:Casimir2Pd1}), we plot the dimensionless combination $
c_1(g) = -  L\,{\mathcal E}_0$ as a function of the (also dimensionless)
variable $g$, for the massless case, and for different values of $M \equiv m
L$, corresponding to the massive case (where there is no closed
expression for the energy).
Note that there is a maximum at $g=2$. That value, as explained in the next
section, corresponds to exact bag boundary conditions, where the boundaries
are more effective and as a consequence the energy reaches it maximum
possible value: 
\begin{equation}
	\big[{\mathcal E}_0\big]_{g=2} \;=\; - \frac{\pi}{24 L}\; 
\end{equation}
Note that in the approach of reference~\cite{Sundberg:2003tc}, bag
conditions correspond to an infinite coupling constant. The difference is
explained, of course, by the different meaning of the coupling constants in
both cases. For example, the propagator in our approach includes
tadpole-like propagation between the wall and itself, which naturally
change the effective value of the coupling constant. 
Nevertheless, our $g=2$ agrees with $\lambda \to \infty$. 

A fuller explanation of this fact is given in section~\ref{sec:bag}, where
we derive the form of the fermion propagator in the presence of the
defects, studying its behaviour close to the boundary.
\begin{figure}[!ht]
\begin{center}
\begin{picture}(0,0)%
\includegraphics{f2PD1mEvsg.pstex}%
\end{picture}%
\setlength{\unitlength}{4144sp}%
\begingroup\makeatletter\ifx\SetFigFont\undefined%
\gdef\SetFigFont#1#2#3#4#5{%
  \reset@font\fontsize{#1}{#2pt}%
  \fontfamily{#3}\fontseries{#4}\fontshape{#5}%
  \selectfont}%
\fi\endgroup%
\begin{picture}(4662,3001)(1414,-3437)
\put(1441,-691){\makebox(0,0)[lb]{\smash{{\SetFigFont{12}{14.4}{\rmdefault}{\mddefault}{\updefault}{\color[rgb]{0,0,0}$c_1(g)$}%
}}}}
\end{picture}%
\caption{$c_1(g)$ as a function of $g$, for different values of $M = m L$.}\label{fg:Casimir2Pd1}
\end{center}
\end{figure}
As already mentioned, for the massive case the corresponding integral
cannot be performed exactly. In Figure 2 we plot the result of numerically
evaluating that integral for $c_1(g, m L) \equiv |{\mathcal E}_0| \,L$ as a
function of $M = m L$, for different values of $g$.
\begin{figure}[!ht]
\begin{center}
\begin{picture}(0,0)%
\includegraphics{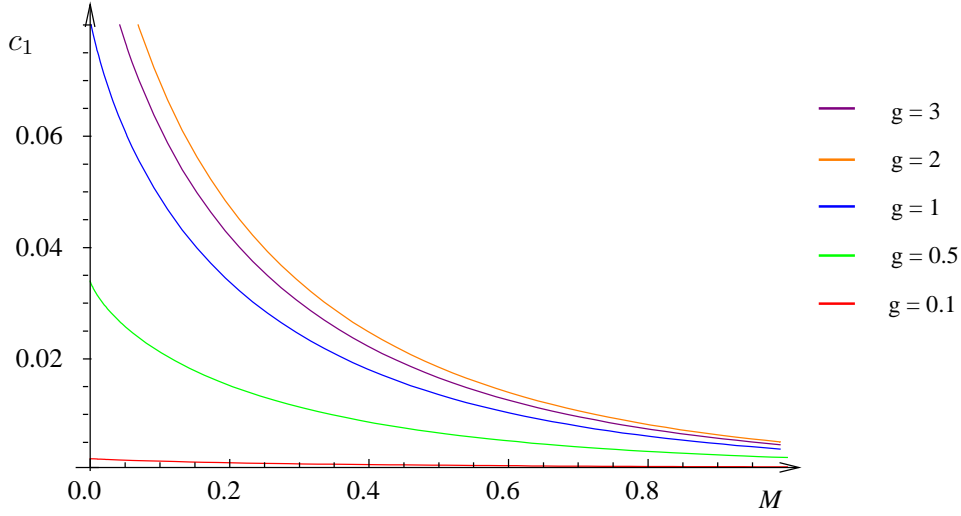}%
\end{picture}%
\setlength{\unitlength}{4144sp}%
\begingroup\makeatletter\ifx\SetFigFont\undefined%
\gdef\SetFigFont#1#2#3#4#5{%
  \reset@font\fontsize{#1}{#2pt}%
  \fontfamily{#3}\fontseries{#4}\fontshape{#5}%
  \selectfont}%
\fi\endgroup%
\begin{picture}(5683,3053)(628,-2911)
\put(643,-161){\makebox(0,0)[lb]{\smash{{\SetFigFont{12}{14.4}{\rmdefault}{\mddefault}{\updefault}{\color[rgb]{0,0,0}$c_1$}%
}}}}
\end{picture}%
\caption{$c_1$ as a function of $M \equiv m L$; $D= 1$.}\label{fg:f2D1mEvsM.jpg}
\end{center}
\end{figure}

In $3$ spatial dimensions, each block in $\widetilde{\mathcal M}$ is a $4
\times 4$ matrix. For a massive field, and using the notation $\omega(q)
\equiv\sqrt{m^2 + q^2}$ we find that:
\begin{eqnarray}
\widetilde{\mathcal K}^{-1}_{00}\widetilde{\mathcal
K}_{01}\widetilde{\mathcal K}^{-1}_{11}\widetilde{\mathcal K}_{10}&=&
\frac{4 g^2 e^{-2\omega(q)L}}{\big[(4\,+\,g^2)\,\omega(q)\,+\,4\,g\,m\big]^2}
\Big[ - 2\,q^2 \,{\mathcal I}\,+\, g\,q^2\,\gamma_D\nonumber\\
&+&i\,(2\,m\,+\,g\,\omega(q))\,\not \! q \,+\,i\, 
(2\,\omega(q)\,+ g\,m)\,{\not \! q}\,\gamma_D \Big]\;,
\end{eqnarray}
with $q = (p_0, p_1, p_2)$, $p_D = p_3$ and  $\gamma_D = \gamma_3$.

The exact eigenvalues of $\widetilde{\mathcal M}$ can be found after some
algebra, the result being:
\begin{eqnarray}
\lambda_1 &=& 1 \nonumber\\ 
\lambda_2 &=& 1
+\,\frac{16\,g^2\,\,q^2}{\big[(4\,+\,g^2)\,\omega(q)\,+\,4\,g\,m \big]^2}\,e^{-2
\omega(q) L} \;,
\end{eqnarray}
each one with multiplicity equal to two.  Then, after integrating out
the angular variables, the Casimir energy per unit area may be written as
follows: 
\begin{equation}
{\mathcal E}_0(g, m, L) \,=\,- \frac{c_3(g, m L)}{L^3}
\end{equation}
where
\begin{equation}
c_3(g,M) \,\equiv \,\frac{1}{\pi^2}\int_{0}^{\infty}dr\,r^2\,
\ln\Big\{1\,+\,\frac{16\,g^2\,r^2\,e^{-2\sqrt{r^2\,+\,M^2}}}{\big[4 g
M\,+\,(4\,+\,g^2)\,\sqrt{r^2\,+\,M^2}\big]^2}\Big\}\;,
\end{equation}
which converges in both the IR and UV regions.

It is convenient to introduce first the result corresponding to the case
$m=0$ and $g=2$, where the energy reaches a maximum:
\begin{equation}
{\mathcal E}_0(2,0,L) \,=\, - \frac{7 \pi^2}{2880 \,L^3} \;,
\end{equation}
in agreement with~\cite{Johnson:1975zp} (see also~\cite{Queiroz:2004wi}).

\begin{figure}[!ht]
\begin{center}
\begin{picture}(0,0)%
\includegraphics{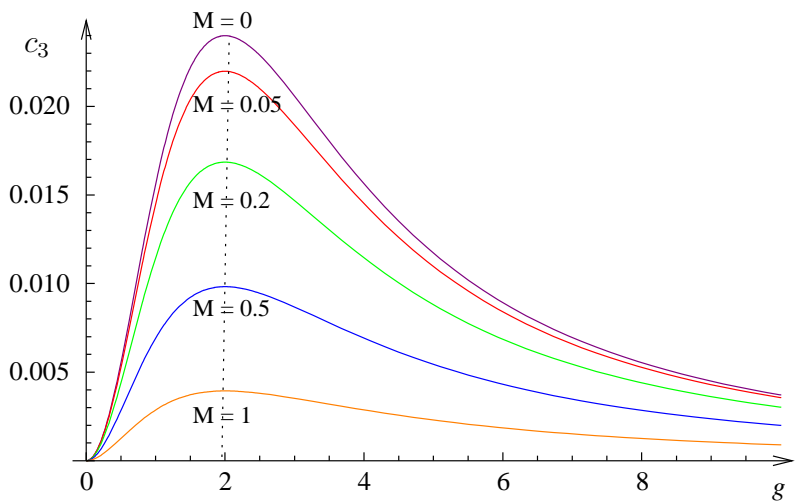}%
\end{picture}%
\setlength{\unitlength}{3947sp}%
\begingroup\makeatletter\ifx\SetFigFont\undefined%
\gdef\SetFigFont#1#2#3#4#5{%
  \reset@font\fontsize{#1}{#2pt}%
  \fontfamily{#3}\fontseries{#4}\fontshape{#5}%
  \selectfont}%
\fi\endgroup%
\begin{picture}(4908,3091)(346,-3107)
\put(451,-286){\makebox(0,0)[lb]{\smash{{\SetFigFont{12}{14.4}{\rmdefault}{\mddefault}{\updefault}{\color[rgb]{0,0,0}$c_3
$}%
}}}}
\end{picture}%
\caption{$c_3$ as a function of $g$, for different values of $M$.}\label{fg:Evsg.jpg}
\end{center}
\end{figure}

When $m \neq 0$, and introducing $M = m\,L \neq 0$, 
\begin{equation}
L^3\,{\mathcal E}_0 \;= \; \frac{1}{\pi^2}\int_{0}^{\infty}dr\;r^2\,
\ln\;\Big[1\,+\,\frac{16\,g^2\,r^2\,e^{-2\sqrt{r^2\,+\,M^2}}}{[\,4\,g\,M\,+\,(4\,+\,g^2)\,\sqrt{r^2\,+\,M^2}\;]^2}\Big]
\;.
\end{equation}
This integral cannot be found analytically, but it has a convenient form
for performing it numerically. In~\ref{fg:Evsg.jpg}, we plot 
$c_3(g,M) = L^3\,|{\mathcal E}_0|$ as a function of $g$ for
different values of $M = m\,L$ between $0.1$ and $1$. We see the expected
decreasing behaviour of the energy when $m$ grows, regardless of the value
of $g$.  This happens because the fermionic field propagator has a faster
decay with distance the larger is the mass, reducing the interaction energy
between them.
On the other hand, the energy always has a maximum when $g = 2$, for any
value of the mass. As we mentioned for the $D=1$ case, this comes from the
bag boundary conditions, which are met precisely at this value.
In figure~\ref{fg:EvsM.jpg}, we present the complementary view, by plotting
$c_3$ as a function of $M$, for different values of $g$ between $0.1$ to $3$. 
It can be seen that the energy becomes negligible when $M \simeq 2$ which
means that the distance between the plates is of the order of the decay
length of the propagator. 
\begin{figure}[!ht]
\begin{center}
\begin{picture}(0,0)%
\includegraphics{f2d3M.pstex}%
\end{picture}%
\setlength{\unitlength}{4144sp}%
\begingroup\makeatletter\ifx\SetFigFont\undefined%
\gdef\SetFigFont#1#2#3#4#5{%
  \reset@font\fontsize{#1}{#2pt}%
  \fontfamily{#3}\fontseries{#4}\fontshape{#5}%
  \selectfont}%
\fi\endgroup%
\begin{picture}(5390,2870)(414,-2551)
\put(429,125){\makebox(0,0)[lb]{\smash{{\SetFigFont{12}{14.4}{\rmdefault}{\mddefault}{\updefault}{\color[rgb]{0,0,0}$c_3$}%
}}}}
\end{picture}%
\caption{$c_3$ as a function of $M = m L$, for different
values of $g$.}\label{fg:EvsM.jpg}
\end{center}
\end{figure}

\section{The $\delta$-interaction and bag boundary conditions}\label{sec:bag}
Bag boundary conditions are usually formulated in terms of the fermionic
propagator \mbox{$G(x,x')=\langle\psi(x)\bar\psi(x')\rangle$}, in the presence of
boundaries.  For the case of just one defect located at $x_d=0$, 
one says that it imposes bag conditions on its two, `right' and `left'
faces, when:
\begin{equation}
\lim_{x_D \to 0\pm} \,( 1 \,\pm\,\gamma_D)\,   G(x,x') \;=\; 0 \;,	
\end{equation}
respectively. They imply that the vacuum expectation value of the normal component of the 
fermionic current vanishes when approaching the wall from each side.
Similar conditions may be imposed, of course, on more than one wall.

Following an analogous procedure to the one use for the calculation of
${\mathcal Z}$, but now for the propagator:
\begin{equation}
	\langle\psi(x)\bar\psi(x')\rangle \;=\;\frac{\int {\mathcal D}\psi
	{\mathcal D}{\bar\psi} \, \psi(x) {\bar\psi}(x') \,
	e^{-S}}{\int {\mathcal D}\psi {\mathcal D}{\bar\psi} \,	e^{-S}} 
\end{equation}
we obtain:
\begin{equation}
G(x,x') = G^{(0)}(x,x')\,+\,T(x,x')\;,
\end{equation}
where:
\begin{eqnarray}
	T(x,x')&=&-g\,\sum_{\alpha ,\beta}\int \frac{d^dq}{(2\pi)^d}\,e^{-iq(y-y')} 
	{\tilde G^{(0)}}(q,x_D,a_\alpha)\nonumber\\
	&\times& [{\tilde K}^{-1}]_{\alpha\,\beta}(q)\,\tilde{G^{(0)}}(q,a_\beta,x'_D)\,.
\end{eqnarray}
with
\begin{equation}
	{\tilde G^{(0)}}(q,x_D,x'_D)\,=\,
	\int \frac{dp_D}{2\pi}\,e^{-ip_D(x_D - x_D')}\, G^{(0)}(x,x') \;.
\end{equation}
The matrix elements of the inverse of ${\tilde K}$ may be written in terms
of the matrix elements of ${\tilde K}$:
\begin{equation}
	\tilde{\mathcal K}^{-1}\,=\,
	\left(
	\begin{array}{cc}
		{C_0^{-1}}&{- [\tilde{\mathcal K}_{00}]^{-1} \tilde{\mathcal K}_{01} C_1^{-1}}\\
\\
{-C_1^{-1} \tilde{\mathcal K}_{10} [\tilde{\mathcal K}_{00}]^{-1}} & {C_1^{-1}}
\end{array}
\right) \;,
\end{equation}
where $[\tilde{\mathcal K}_{00}]^{-1}$ and $[\tilde{\mathcal K}_{11}]^{-1}$
denote the inverses of the respective matrix elements (not to be confused
with the matrix elements of the inverse).
We have used the definitions $C_0 \equiv \tilde{\mathcal
K}_{00}-\tilde{\mathcal K}_{01}[\tilde{\mathcal K}_{11}]^{-1}
\tilde{\mathcal K}_{10}$  and  
$C_1 \equiv \tilde{\mathcal K}_{11}-\tilde{\mathcal K}_{10}[\tilde{\mathcal
K}_{00}]^{-1} \tilde{\mathcal K}_{01}$. 

Let us apply the formulae above, for the sake of simplicity, to the case of massless fermions and only one
defect in $D = 1$, we see that, when  $x_1 \rightarrow 0^+$ and $x'_1 > 0$ we obtain:
$$
(1 \pm \gamma_1) \big[\tilde{G}(q,0,x'_1)\,+\,\tilde{G}(q,0,x'_1)\,\big] \,=\,
e^{-\,\vert q \vert \vert {x'}_1\vert}\,\big(\frac{2\,g}{4 + g^2}\;\mp\,
\frac{1}{2}\big)\
$$
\begin{equation}
	\times \Big[\mathcal{I}\;\mp\,i\,{\rm
	sg}(q)\,\gamma_0\,\pm\,\gamma_1\,+\,i\,{\rm sg}(q)\,
	\gamma_0\gamma_1\Big]\,\;,
\end{equation}
where ${\rm sg}$ is the sign function.
Since we are interested in the $+$ sign when approaching the defect from
the right, we see that the bag condition is fulfilled when
\begin{equation}
\frac{2\,g}{4\,+\,g^2}\,=\,\frac{1}{2}
\end{equation}
thus, $g=2$.

On the other hand, when one approaches the defect from the left, one also
obtains $g=2$ to satisfy the corresponding bag condition. An entirely
analogous derivation yields the same value for $g$ when the fermions are
massive, or when more dimensions are considered. For example, in $2+1$
dimensions, and for $m=0$ the propagator reduces to the one
of~\cite{Fosco:2004cn}.

\section{Conclusions}\label{sec:conc}
We derived general expressions for the vacuum energy in the presence of $N$
`defects', whose role is to impose {\em imperfect\/} bag-like boundary
conditions on parallel planes. Standard bag conditions may be obtained as a
particular case. For the case of two plates, we obtain more explicit
formulae, which yield the Casimir energy as an integral over a single
variable. That integral is both UV and IR finite.

We show that, at any given distance between the plates, the energy 
reaches a maximum precisely when the bag condition is satisfied. 

Our way of implementing the calculation relies on the calculation of the
exact fermion propagator in the presence of the defects. This propagator is
a perfectly well-defined object, in spite of the fact that the eigensystem
for the associated Dirac Hamiltonian is not well-defined. This ambiguity
has the consequence of allowing for different parametrizations of the
strength of the coupling between the Dirac field and the mirrors. In
particular, our method produces a fermionic propagator which
satisfies bag boundary conditions when the coupling constant $g=2$. 
That value corresponds, in the setting of reference~\cite{Sundberg:2003tc} to an infinite
coupling constant.
\newpage
\section*{Acknowledgments}
C.~D.~F.~and E.~L.~L.~acknowledge support from CONICET, ANPCyT and UNCuyo.

\end{document}